# Operationalistic Orthogonality Condition for Single-Mode Biphotons (Qutrits)

A. A. Zhukov, G. A. Maslennikov, and M. V. Chekhova

*Moscow State University, Vorob'evy gory, Moscow, 119899 Russia*
*e-mail: postmast@qopt.phys.msu.su*

An arbitrary polarization state of a single-mode biphoton is considered. The operationalistic criterion is formulated for the orthogonality of these states. It can be used to separate a biphoton with an arbitrary degree of polarization from a set of biphotons orthogonal to it. This is necessary for the implementation of quantum cryptography protocol based on the three-level systems. The experimental test of this criterion amounts to the observation of the anticorrelation effect for a biphoton with an arbitrary polarization state. © 2002 MAIK "Nauka/Interperiodica".

In recent years, considerable interest has been shown in multilevel systems with dimensionality higher than two, because they provide a way for more dense data recording, as compared to the traditional two-level systems. This is particularly important for quantum cryptography, because it enables one to increase the data-exchange rate [1] and enhance security against eavesdropping attacks of a certain class [2]. However, the transition to higher-dimensionality systems inevitably gives rise to experimental difficulties associated with the implementation and the adequate measurement of parameters. The solution of these problems requires the design of a data output device, error-correction protocols, repeaters, and other quantum communication devices. After the two-level system, the three-level system is the simplest. Its state in quantum information is called "qutrit" by analogy with qubit. The wave function of an arbitrary three-level system can be written as

$$\Psi = c_1|1\rangle + c_2|2\rangle + c_3|3\rangle, \tag{1}$$

where $|1\rangle$, $|2\rangle$, and $|3\rangle$ are the orthogonal basis states. The complex coefficients $c_i$ are called the amplitudes of basis states $|i\rangle$ and related to each other by the normalization condition

$$\sum_{i=1}^{3} |c_i|^2 = 1. \tag{2}$$

At present, the use of qutrits in quantum information is not an exotic thing. For example, the authors of [3] have proposed a quantum cryptography protocol based on the three-level systems, and the interferometric method of preparing multilevel systems was considered in [1]. The theoretical analysis of state restoration from the measured quantities was carried out for an arbitrary multilevel system in [4].

Biphoton fields have been used in many experiments of quantum optics practically since its inception. These fields are fluxes of photon pairs strongly correlated in coordinate and time of their creation. In the great majority of experiments, spontaneous parametric down-conversion (SPDC) is used as a source of biphoton field. It will be shown below that, in the case where a photon pair is emitted into the same spatial and frequency mode, the state polarization characteristics of the biphoton allow it to be considered as a qutrit [5]. The use of single-mode biphotons as qutrits in quantum communication protocols, e.g., in quantum cryptography poses the problem of separating a certain biphoton from the subset of biphotons with polarizations orthogonal to the polarization of the former. In this work, the operationalistic biphoton orthogonality condition is formulated for an arbitrary degree of polarization, and the experimental scheme is proposed that allows unambiguous separation of a certain biphoton polarization state from a set of states orthogonal to it. This is a fundamental problem, and its solution can be used, e.g., in the practical implementation of quantum cryptography protocol for the three-level systems.

**Biphotons and qutrits.** The use of the polarization states of single-mode biphotons for data recording was proposed in [5], and the polarization characteristics of these fields were discussed in [6, 7]. A pure polarization state of a biphoton in the collinear frequency-degenerate regime can be written as [5]

$$|\Psi\rangle = c_1|2, 0\rangle + c_2|1, 1\rangle + c_3|0, 2\rangle. \tag{3}$$

Here, $|n, m\rangle$ denote the state with $n$ photons in the horizontal ($H$) polarization mode and $m$ photons in the vertical ($V$) mode ($n + m = 2$) and $c_i = |c_i|\exp(i\phi_i)$ is the

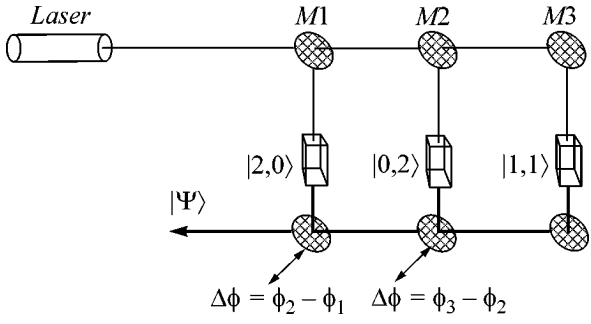

**Fig. 1.** Scheme of a nonlinear three-arm Mach–Zehnder interferometer. Nonlinear crystals oriented so as to produce the appropriate state are placed in each of the arms. By introducing (with the use of mirrors) phase difference between the states, one can change the phases of coefficients $c_i$ in the desired way and, by varying the pump power, one can change their amplitudes in a desired way. Mirrors are denoted by $M1$, $M2$, and $M3$.

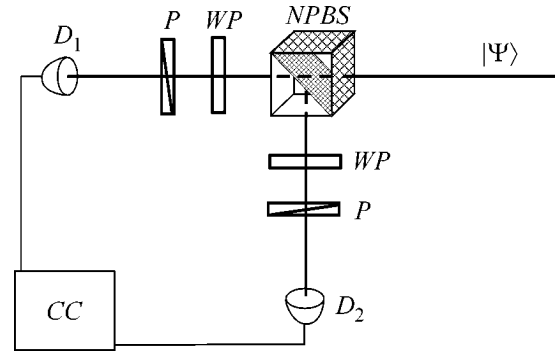

**Fig. 2.** Scheme for measuring arbitrary polarization state of a biphoton field. *NPBS* is the nonpolarizing beam splitter, *WP* is a set of wave plates, *P* is the analyzer, *D* is a single-photon detector, and *CC* is the coincidence counting scheme.

complex amplitude of the probability of finding biphoton in the corresponding state. The states $|2, 0\rangle$, $|1, 1\rangle$, and $|0, 2\rangle$ constitute the orthogonal basis set in so-called *HV* representation. By analogy with Eq. (1), this state can be used for the ternary information coding. To visualize the polarization states of a biphoton light, it is convenient to use the Poincaré sphere [7]. One can show that state vector (3) can be represented as

$$|\Psi\rangle = \frac{a^\dagger(\vartheta, \varphi)a^\dagger(\vartheta', \varphi')|vac\rangle}{\|a^\dagger(\vartheta, \varphi)a^\dagger(\vartheta', \varphi')|vac\rangle\|}, \quad (3a)$$

where $a^\dagger(\vartheta, \varphi)$ and $a^\dagger(\vartheta', \varphi')$ are the operators of photon creation and annihilation in an arbitrary polarization mode, e.g.,

$$a^\dagger(\vartheta, \varphi) = \cos\frac{\vartheta}{2}a_H^\dagger + e^{i\varphi}\sin\frac{\vartheta}{2}a_V^\dagger,$$

where $a_{H,V}^\dagger$ are the operators of photon creation in the linear polarization modes $H$ and $V$, and $\varphi, \varphi' \in [0, 2\pi]$ and $\vartheta, \vartheta' \in [0, \pi]$ are, respectively, the azimuthal and polar angles on the Poincaré sphere. In this case,

$$\varphi, \varphi' = \frac{\varphi_3}{2} \pm \frac{1}{2}\arccos\left[\frac{|c_2|^2}{2|c_1||c_3|}\right.$$
$$\left. - \sqrt{1 + \frac{|c_2|^4}{4|c_1|^2|c_3|^2} - \frac{|c_2|^2}{|c_1||c_3|}\cos(2\varphi_2 - \varphi_3)}\right],$$

$$\theta, \theta' = \arccos\{[|c_1|^2 - |c_3|^2$$
$$\pm 2\sqrt{[|c_2|^2 - |c_1||c_3|\cos(2\varphi - \phi_3)]^2 - |c_1|^2|c_3|^2}]$$
$$\times [1 + |c_2|^2 - 2|c_1||c_3|\cos(2\varphi - \phi_3)]^{-1}\},$$

where the unprimed and primed quantities relate, respectively, to the first and the second photon of the pair. It is also convenient to introduce the degree of biphoton polarization. This quantity was calculated in [6] and, in the new representation, takes the form

$$P = \frac{2\cos(\alpha/2)}{1 + \cos^2(\alpha/2)},$$

where $\alpha$ is the angle at which the pair of points mapping biphoton onto the Poincaré sphere is seen from its center.

The generation of a biphoton field in an arbitrary polarization state with given coefficients $c_i$ can be accomplished in a three-arm interferometer of the Mach–Zehnder type with three nonlinear crystals. Each of them produces one of the basis states in the corresponding arm to prepare a linear superposition of type (3) at the output. The amplitudes and phases of the complex coefficients $c_i$ can be varied in a desired way in each of the arms (Fig. 1). In particular, all states used in the quantum cryptography protocol proposed in [3] can be prepared in the interferometer of this type.

To adequately measure the polarization characteristics of single-mode biphoton fields, the Braun–Twiss scheme is used with arbitrary polarizing filters in the arms (Fig. 2). Each filter includes a polarization transformer and a linear polarization analyzer to separate a certain polarization state. Upon measuring a certain set of fourth-order field moments in this scheme, one can determine the real and imaginary parts of coefficients $c_i$ by varying the transformer characteristics. For some pure states $|\Psi\rangle$ this was done in [8]. However, this scheme can be used to measure an arbitrary polarization state of biphoton field in the *HV* basis. We will refer to the measuring scheme as "tuned" to the particular polarization biphoton state (3a) if the polarization state with parameters $(\vartheta, \varphi)$ is separated in one channel and the state with parameters $(\vartheta', \varphi')$ in the other.

**Orthogonality of single-mode biphotons.** Assume that the polarization states corresponding to modes $a_1$ and $b_2$ (letters denote the polarization states and index denotes the spatial mode) are separated in channels 1 and 2 of a device illustrated in Fig. 2. Let us write the orthogonality condition for a certain input state $|\Psi_{cd}\rangle$ and the state $|\Psi_{ab}\rangle$ to which the detector device is tuned. Let

$$|\Psi_{ab}\rangle = \frac{a^\dagger b^\dagger |\text{vac}\rangle}{\|a^\dagger b^\dagger |\text{vac}\rangle\|}, \quad |\Psi_{cd}\rangle = \frac{c^\dagger d^\dagger |\text{vac}\rangle}{\|c^\dagger d^\dagger |\text{vac}\rangle\|},$$

where $a^\dagger$, $b^\dagger$, $c^\dagger$, and $d^\dagger$ are the creation operators for the polarization modes $a$, $b$, $c$, and $d$, respectively. We emphasize that, in the general case, the modes $a$, $b$, $c$, and $d$ are not mutually orthogonal. The condition for the orthogonality of the input biphoton and the biphoton to which the device is tuned has the form

$$\langle \Psi_{ab} | \Psi_{cd} \rangle = 0$$

or, equivalently,

$$\langle \text{vac}|abc^\dagger d^\dagger|\text{vac}\rangle = 0. \qquad (4)$$

Since the creation and annihilation operators before and after a 50% beam splitter are related to each other as $a^\dagger = \frac{1}{\sqrt{2}}(a_1^\dagger + i a_2^\dagger)$, Eq. (4) can be rewritten as

$$\langle \text{vac}|(a_1 - ia_2)(b_1 - ib_2)(c_1^\dagger + ic_2^\dagger)(d_1^\dagger + id_2^\dagger)|\text{vac}\rangle = 0. \qquad (5)$$

Neglecting the terms $\langle \text{vac}|a_i b_i c_i^\dagger d_i^\dagger|\text{vac}\rangle$, which are equal to zero according to Eq. (4), and the terms of the form $\langle \text{vac}|a_i b_i c_j^\dagger d_j^\dagger|\text{vac}\rangle$, which are also equal to zero because they correspond to the creation of a photon pair in the mode $j$ and annihilation of a photon pair in the mode $i$, etc., one gets

$$\langle \text{vac}|a_1 b_2(c_1^\dagger d_2^\dagger + c_2^\dagger d_1^\dagger) + a_2 b_1(c_1^\dagger d_2^\dagger + c_2^\dagger d_1^\dagger)|\text{vac}\rangle = 0. \qquad (6)$$

Note that

$$\langle \text{vac}|a_1 b_2(c_1^\dagger d_2^\dagger + c_2^\dagger d_1^\dagger)|\text{vac}\rangle$$
$$= \langle |a_1 c_1^\dagger\rangle\langle |b_2 d_2^\dagger\rangle + \langle |a_1 d_1^\dagger\rangle\langle |b_2 c_2^\dagger\rangle$$
$$= \langle |ac^\dagger\rangle\langle |bd^\dagger\rangle + \langle |ad^\dagger\rangle\langle |bc^\dagger\rangle,$$

$$\langle \text{vac}|a_2 b_1(c_1^\dagger d_2^\dagger + c_2^\dagger d_1^\dagger)|\text{vac}\rangle$$
$$= \langle |a_2 c_2^\dagger\rangle\langle |b_1 d_1^\dagger\rangle + \langle |a_2 d_2^\dagger\rangle\langle |b_1 c_1^\dagger\rangle$$
$$= \langle |ac^\dagger\rangle\langle |bd^\dagger\rangle + \langle |ad^\dagger\rangle\langle |bc^\dagger\rangle;$$

i.e., the mean values of both terms in Eq. (6) are equal, because they differ only in the spatial indices and, hence,

$$\langle \text{vac}|a_1 b_2(c_1^\dagger d_2^\dagger + c_2^\dagger d_1^\dagger)|\text{vac}\rangle = 0. \qquad (7)$$

Note that the state vectors of the form $a_i b_j c_i^\dagger d_j^\dagger|\text{vac}\rangle$ in Eq. (7) contain two creation and two annihilation operators; hence, their sum is a vacuum state multiplied by a numerical factor. It follows from Eq. (7) that this factor is zero. Therefore, the orthogonality condition for the biphotons $|\Psi_{ab}\rangle$ and $|\Psi_{cd}\rangle$ finally takes the form

$$a_1 b_2(c_1^\dagger d_2^\dagger + c_2^\dagger d_1^\dagger)|\text{vac}\rangle = 0. \qquad (8)$$

After the beam splitter, the input state $|\Psi_{cd}\rangle$ becomes

$$|\Psi'_{cd}\rangle = \frac{i}{2\|c^+ d^+ |\text{vac}\rangle\|}$$
$$\times \{c_1^\dagger d_2^\dagger + d_1^\dagger c_2^\dagger + c_1^\dagger d_1^\dagger + c_2^\dagger d_2^\dagger\}|\text{vac}\rangle.$$

The last two terms make no contribution to the coincidences, because they correspond to the situation where both photons are led to the same photodetector. For this reason, the coincidence counting rate is determined by the second-order correlation function

$$G^{(2)} = \frac{1}{4\|c^+ d^+ |\text{vac}\rangle\|^2}$$
$$\times \langle \text{vac}|\{c_1 d_2 + d_1 c_2\} a_1^\dagger b_2^\dagger a_1 b_2 \{c_1^\dagger d_2^\dagger - d_1^\dagger c_2^\dagger\}|\text{vac}\rangle. \qquad (9)$$

The absence of photocount coincidences for detectors $D_1$ and $D_2$ is equivalent to the zero value of the correlator in Eq. (9). The condition for the absence of coincidences can be written as

$$\langle \text{vac}|\{c_1 d_2 + d_1 c_2\} a_1^\dagger b_2^\dagger a_1 b_2 \{c_1^\dagger d_2^\dagger + d_1^\dagger c_2^\dagger\}|\text{vac}\rangle = 0, \qquad (10)$$

which is equivalent to condition (8).

Therefore, the condition for the orthogonality of two biphotons is equivalent to the condition for the absence of coincidences in the scheme in Fig. 2, provided that one biphoton is fed into the input, while the device is tuned to the second biphoton. This procedure can be regarded as the projection of one polarization state onto the other, with the number of photocount coincidences playing the role of an observable quantity. In the case that the second state is orthogonal to the initial state, the coincidence counting rate should drop to the level of accidental coincidences. It should also be noted that the rate of single photocounts in both detectors will be, generally, nonzero upon changing the characteristics of polarization transformers, e.g., upon analyzer rotation. In this case, the character of changing the number of single photocounts takes the form of interference pattern with the visibility equal to the degree of polarization of a measured biphoton [9].

The experimental data [5] presented in the table illustrate the biphoton orthogonality criterion. For example, if the recording scheme is tuned to the $|H, V\rangle$ state, while the input state is $|R, L\rangle$ (a pair of right- and Experimental dependence of the coincidence counting rate on the input polarization state of a biphoton and the state to which the device is tuned

| Input state | Degree of polarization, $P$ | Detected state | Degree of polarization, $P$ | Coincidence counting rate ($s^{-1}$) |
|---|---|---|---|---|
| $|H, V\rangle$ | 0 | $|H, V\rangle$ | 0 | $4.0 \pm 0.4$ |
| $|R, L\rangle$ | 0 | $|H, V\rangle$ | 0 | $0.5 \pm 0.25$ |
| $|D, \bar{D}\rangle$ | 0 | $|H, V\rangle$ | 0 | $0.25 \pm 0.1$ |
| $|H, V\rangle$ | 0 | $|D, \bar{D}\rangle$ | 0 | $0.25 \pm 0.1$ |
| $|D, \bar{D}\rangle$ | 0 | $|D, \bar{D}\rangle$ | 0 | $3.8 \pm 0.4$ |
| $|H, V\rangle$ | 0 | $|H, H\rangle$ | 1 | $0.15 \pm 0.05$ |
| $|D, \bar{D}\rangle$ | 0 | $|H, H\rangle$ | 1 | $1.9 \pm 0.2$ |

Note: The following notation is used for the polarization modes: $H$ is the horizontal direction of mode polarization; $V$ is the vertical direction; $D$, $\bar{D}$ is the linear polarization with angles of $+45°$ and $-45°$ to the vertical direction; and $R$ and $L$ are the right- and left-hand circular polarizations, respectively.

left-hand circularly polarized photons) or $|D\bar{D}\rangle$ (a pair of photons linearly polarized at ±45°) is orthogonal to it, the coincidence counting rate is an order of magnitude lower than for the same input state $|H, V\rangle$. Likewise, a low coincidence counting rate is observed in the case where the input state is $|H, V\rangle$, while the device is tuned to the orthogonal state $|H, H\rangle$. At the same time, if the device is tuned to the $|H, H\rangle$ state and $|D\bar{D}\rangle$ is the input state, whose projection onto the $|H, H\rangle$ equals $1/\sqrt{2}$, the coincidence counting rate is half as high as for the case where the device is tuned to the input state.

Thus, the biphoton orthogonality criterion suggested in this work allows one to unambiguously separate a biphoton in an arbitrary polarization state from a set of biphotons orthogonal to it. The experimental test of this criterion amounts to the observation of anticorrelation [10, 11] for an arbitrary biphoton polarization state, in contrast to works [10, 11], where the directions of photon polarization in a pair were identical, or work [12], where they were mutually orthogonal.

In practice, this criterion can be used for the implementation of a quantum cryptography protocol [3]. The possibility of unambiguously separating the desired biphoton polarization state from a set of states orthogonal to it allows a certain logical value used in the secret key to be assigned to this biphoton with assurance. Nevertheless, this scheme is not free from losses. For instance, both photons in a pair may be led into the same arm after beam splitter and, hence, may make no contribution to the coincidences. These processes alone halve the amount of useful information. Another loss source appears because, despite the fact that the suggested scheme filters out only one biphoton $|\Psi_{cd}\rangle = c^\dagger d^\dagger|vac\rangle/\|c^\dagger d^\dagger|vac\rangle\|$ from a set of biphotons orthogonal to it, the probability that this biphoton will make no contribution at the output is nonzero even if the detectors are ideal. This may occur if a photon in mode $c$ is led to the arm tuned to mode $d$, and vice versa. Clearly, since the modes $c$ and $d$ are generally different, each of these photons may not be detected separately and, as a result, no coincidence will occur.

We are grateful to S.P. Kulik and P.A. Prudkovskiĭ for discussions. This work was supported by the Russian Foundation for Basic Research (project nos. 02-02-16664, 00-15-96541) and INTAS (grant no. 01-2122).

*Translated by V. Sakun*